\documentclass[10pt,twocolumn,letterpaper]{article}

\usepackage{iccv}
\usepackage{times}
\usepackage{epsfig}
\usepackage{graphicx}
\usepackage{graphics}

\usepackage{caption}
\usepackage{subcaption}
\usepackage{amsmath}
\usepackage{amssymb}

\usepackage{CJKutf8}

\usepackage{arydshln}
\usepackage{booktabs}
\usepackage{enumitem}
\usepackage{balance}
\usepackage{combelow}
\usepackage{tabularx}
\usepackage{cite}

\usepackage{wrapfig}
\usepackage[stable]{footmisc}

\usepackage{multirow}
\usepackage{makecell}
\usepackage{mathtools}
\usepackage{algorithm}
\usepackage{algorithmicx}
\usepackage{eqparbox,array}

\usepackage{xcolor,colortbl}

\usepackage{subcaption}
\usepackage{caption}
\usepackage{kotex}
\usepackage{verbatim}

\usepackage[accsupp]{axessibility} 
\definecolor{Gray}{gray}{0.85}
\definecolor{darkgreen}{rgb}{0.0, 0.8, 0.0}
\definecolor{Lightgray}{gray}{0.90} 
\newcolumntype{a}{>{\columncolor{Lightgray}}c}

\newcommand\blfootnote[1]{%
  \begingroup
  \renewcommand\thefootnote{}\footnote{#1}%
  \addtocounter{footnote}{-1}%
  \endgroup
}

\usepackage[pagebackref=true,breaklinks=true,letterpaper=true,colorlinks,bookmarks=false]{hyperref}

\iccvfinalcopy 

\def\iccvPaperID{6967} 

\ificcvfinal\fi

\DeclareMathOperator*{\argmin}{arg\,min}
\begin{document}

\title{Rethinking Deep Image Prior for Denoising}


\author{Yeonsik Jo$^\S$\\
LG AI Research\\
{\tt\small yeonsik.jo@lgresearch.ai}
\and
Se Young Chun\\
ECE, INMC, Seoul National University\\
{\tt\small sychun@snu.ac.kr}
\and
Jonghyun Choi$^{\dagger}$\\
GIST, South Korea\\
{\tt\small jhc@gist.ac.kr}
}

\maketitle
\ificcvfinal\thispagestyle{empty}\fi
\begin{abstract}
Deep image prior (DIP) serves as a good inductive bias for diverse inverse problems.
Among them, denoising is known to be particularly challenging for the DIP due to noise fitting with the requirement of an early stopping.
To address the issue, we first analyze the DIP by the notion of effective degrees of freedom (DF) to monitor the optimization progress and propose a principled stopping criterion before fitting to noise without access of a paired ground truth image for Gaussian noise.
We also propose the `stochastic temporal ensemble (STE)' method for incorporating techniques to further improve DIP's performance for denoising.
We additionally extend our method to Poisson noise.
Our empirical validations show that given a single noisy image, our method denoises the image while preserving rich textual details. Further, our approach outperforms prior arts in LPIPS by large margins with comparable PSNR and SSIM on seven different datasets.\blfootnote{\hspace{-1.8em}$^\S$: work done while with GIST. $^\dagger$: corresponding author. \\
{\bf Code}: {\url{https://github.com/gistvision/DIP-denosing}}}
\end{abstract} 
\vspace{-1em}

\section{Introduction}

Deep neural network has been widely used in many computer vision tasks, yielding significant improvements over conventional approaches since AlexNet~\cite{Alexnet_2012_Alex}.
However, image denoising has been one of the tasks in which conventional methods such as BM3D~\cite{BM3D_2007_Dabov} outperformed many early deep learning based ones~\cite{SDA2010Pascal, MLP_2012_CVPR, DNNAllgaussian2014wang} until DnCNN~\cite{DnCNN_2017_TIP} outperforms it for synthetic Gaussian noise at the expense of massive amount of noiseless and noisy image pairs~\cite{DnCNN_2017_TIP}.

\begin{figure}[h!]
    \begin{subfigure}{0.325\linewidth}
    \caption{Image}
        \includegraphics[width =1\textwidth]{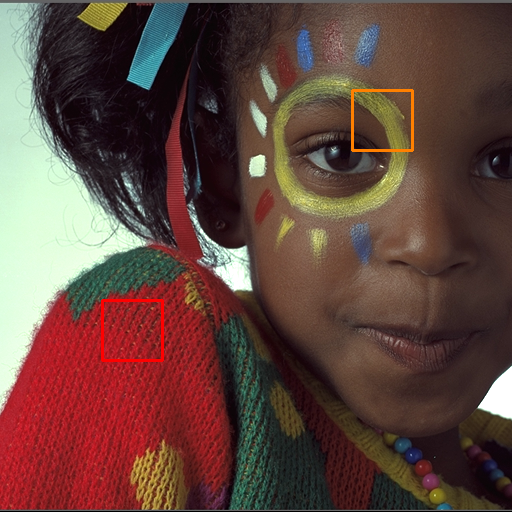}
        \vspace{0.05em}
    \end{subfigure}
    \begin{subfigure}{0.64\linewidth}
    \caption{Comparison on CSet9 dataset}
        \includegraphics[width =1\textwidth]{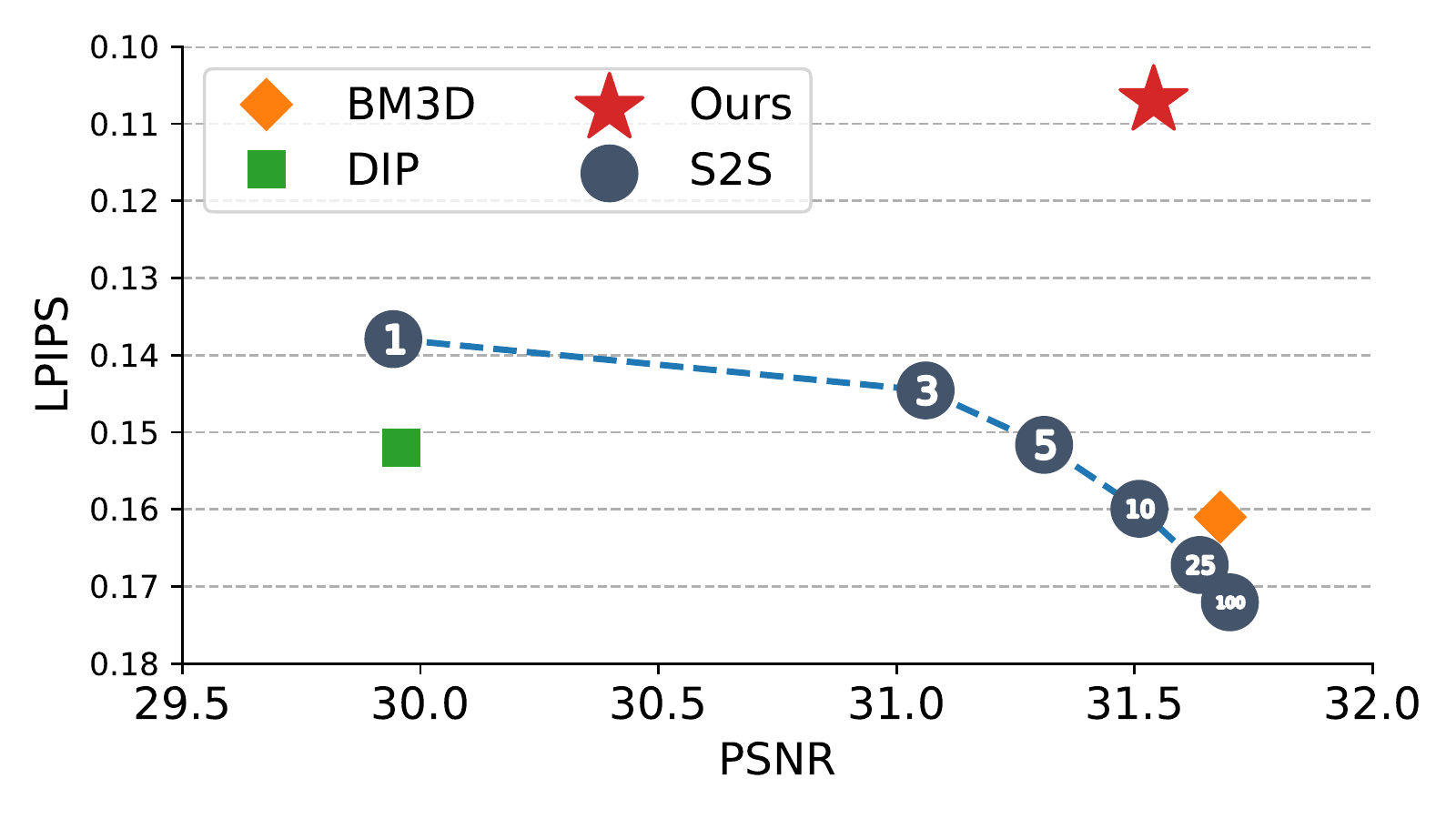} 
    \end{subfigure}
    
    \begin{subfigure}{0.075\textwidth}
    \centering\includegraphics[width =1\textwidth]{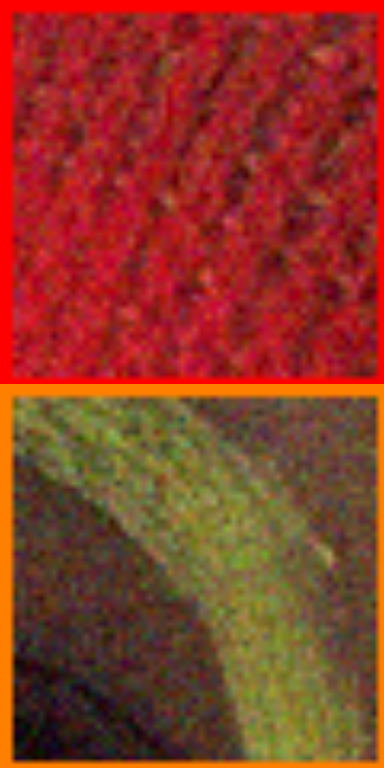}
    \caption*{Noise\\ 20.83/0.56 \centering}
    \end{subfigure}
    \begin{subfigure}{0.075\textwidth}
    \centering\includegraphics[width =1\textwidth]{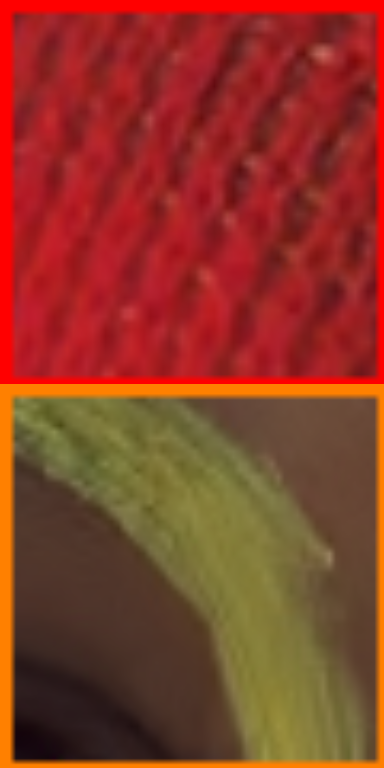}
    \caption*{BM3D\\ \textbf{33.06}/\underline{0.16} \centering}
    \end{subfigure}
    \begin{subfigure}{0.075\textwidth}
    \centering\includegraphics[width =1\textwidth]{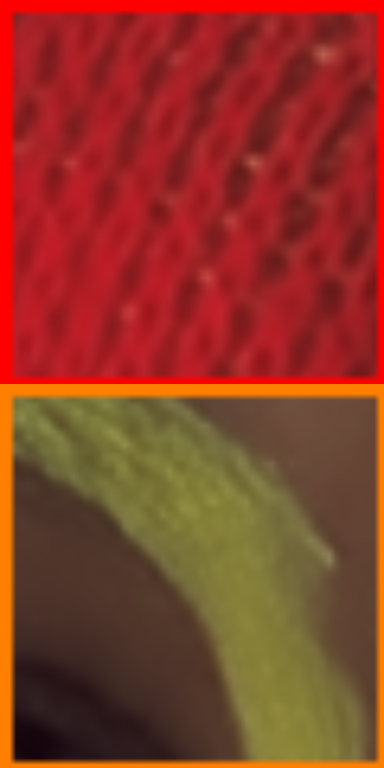}
    \caption*{S2S \\ 32.83/0.18\centering}
    \end{subfigure}
    \begin{subfigure}{0.075\textwidth}
    \centering\includegraphics[width =1\textwidth]{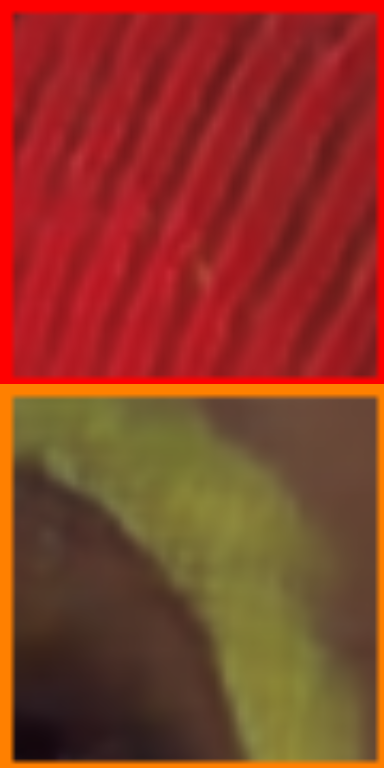}
    \caption*{DIP \\ 30.92/0.18 \centering}
    \end{subfigure}
    \begin{subfigure}{0.075\textwidth}
    \centering\includegraphics[width =1\textwidth]{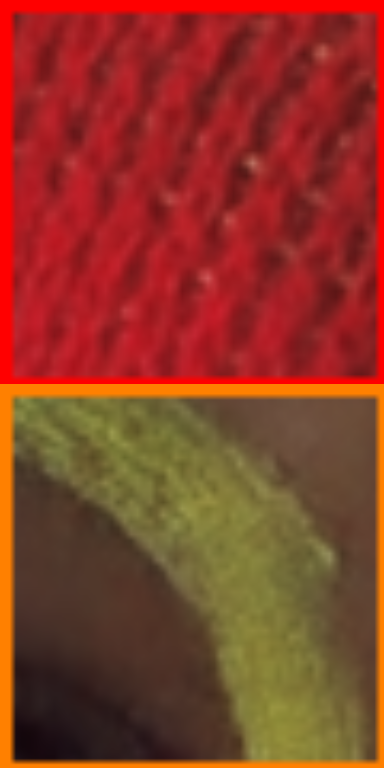}
    \caption*{Ours \\ \underline{32.86}/\textbf{0.14}\centering}
    \end{subfigure}
    \begin{subfigure}{0.075\textwidth}
    \centering\includegraphics[width =1\textwidth]{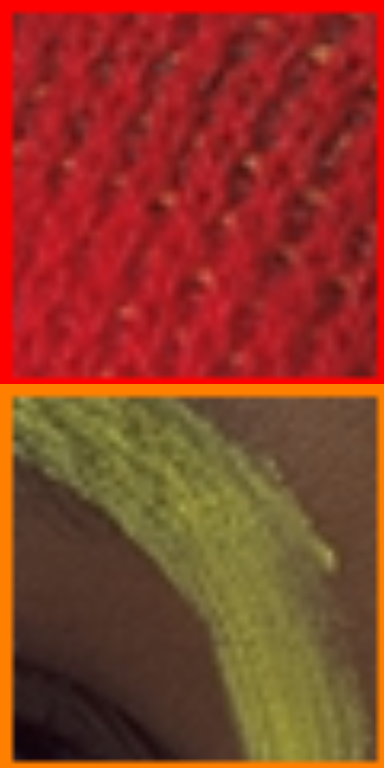}
    \caption*{GT \\ P $\uparrow$ / L$\downarrow$\centering}
    \end{subfigure}\\
\vspace{-0.5em}
\caption{\textbf{Comparison of single image based denoising methods.} `L$\downarrow$' refers to LPIPS and it is lower the better. `P$\uparrow$' refers to PSNR and it is higher the better. Our method denoises an image while preserves rich details; showing the best LPIPS with comparable PSNR to Self2Self (S2S)~\cite{S2S_Quan_2020_CVPR}. Ours shows much better trade-off in PSNR and LPIPS than all other methods including all different ensembling attempts of state of the art (S2S). (Numbers in the circle of S2S denotes number of models in ensemble)}
\vspace{-1.5em}
\label{fig:cmp_image_T1}
\end{figure}

Requiring no clean and/or noisy image pairs, deep image prior (DIP)~\cite{DIP_2018_CVPR,DIP_2020_IJCV} has shown that a randomly initialized network with hour-glass structure acts as a prior for several inverse problems including denoising, super-resolution, and inpainting with a single degraded image.
Although DIP exhibits remarkable performance in these inverse problems, denoising is the particular task that DIP does not perform well, \ie, a single run yields far lower PSNR than BM3D even for synthetic Gaussian noise set-up~\cite{DIP_2018_CVPR,DIP_2020_IJCV}.
Furthermore, for the best performance, one needs to monitor the PSNR (\ie, the ground-truth clean image is required here) and stop the iterations before fitting to noise.
Deep Decoder addresses the issue by proposing a strong structural regularization to allow longer iterations for the inverse problems including denoising~\cite{Deep_heckel_2018_ICLR}. 
However, it yields worse denoising performance than DIP due to low model complexity. 

For better use of DIP for denoising without monitoring PSNR with a clean image, we first analyze the model complexity of the DIP by the notion of effective degrees of freedom (DF)~\cite{Tibshirani2014DoF, df2004Efron, dfinDNN2016UAI}.
Specifically, the DF quantifies the amount of overfitting (\ie, optimism) of a chosen hypothesis (\ie, a trained neural network model) to the given training data~\cite{df2004Efron}. 
In other words, when overfitting occurs, the DF increases.
Therefore, to prevent the overfitting of the DIP network to the noise, we want to suppress the DF over iterations. 
But obtaining DF again requires a clean (ground truth) image.
Fortunately, for the Gaussian noise model, there are approximations for DF without using a clean image; Monte-Carlo divergence approximations in Stein's unbiased risk estimator (SURE) (Eqs.~\ref{eq:sure_def},~\ref{eq:MCdiv2}) ($DF_{MC}$).

Leveraging SURE and improvement techniques in DIP~\cite{DIP_2020_IJCV}, we propose an objective with `stochastic temporal ensembling (STE),' which mimics ensembling of many noise realizations in a single optimization run.
On the proposed objective with the STE, we propose to stop the iteration when the proposed objective function crosses zero.
The proposed method leads to much better solutions than DIP and outperforms prior arts for single image denoising. 
%
In addition, inspired by PURE formulation~\cite{PURE2011Luisier, PGURE2014Montagner}, we extend our objective function to address the Poisson noise.

We empirically validate our method by comparing DIP based prior arts for denoising performance in various metrics that are suggested in the literature~\cite{Gu2019ABR} such as PSNR, SSIM and learned perceptual image patch similarity (LPIPS)~\cite{LPIPS_zhang2018perceptual} on seven different datasets.
LPIPS has been widely used in super resolution literature to complement PSNR, SSIM to measure the recovery power of details~\cite{Ledig2017SRGAN}. 
Since it is challenging for denoiser to suppress noise and preserve details together~\cite{PDTradeoff2018Blau}, we argue that LPIPS is another appropriate metric to evaluate denoisers. 
Note that it has not been widely used in denoising literature yet to analyze the denoising performance. 
Our method not only denoises the images but also preserves rich textual details, outperforming other methods in LPIPS with comparable classic measures including the PSNR and SSIM.

Our contributions are summarized as follows:
\vspace{-1em}
\begin{itemize}
\setlength\itemsep{-0.5em}
    \item Analyzing the DIP for denoising with effective degrees of freedom (DF) of a network and propose a loss based stopping criterion without ground-truth image.
    \item Incorporating noise regularization and exponential moving average by the proposed stochastic temporal ensembling (STE) method.
    \item Diverse evaluation in various metrics such as LPIPS, PSNR and SSIM in seven different datasets.
    \item Extending our method to Poisson noise.
\end{itemize}

\section{Related work}
\label{sec:related}

\subsection{Learning based methods}
Learning-based denoising methods use a large number of clean-noisy image pairs to train a denoiser.
In an early study, a neural network shows decent performance even though the noisy level is unknown, \ie, blind noise setup~\cite{MLP_2012_CVPR}.
Shortly afterwards, however, \cite{DND_Plotz_2017_CVPR} has shown that most of early learning-based studies have often produced worse results than the classical technique such as BM3D~\cite{BM3D_2007_Dabov}.
But recently, DnCNN model with residual learning~\cite{DnCNN_2017_TIP} outperforms BM3D.
Then, several works are proposed to improve the computational efficiency, IRCNN~\cite{IRCNNzhang_2017_learning} uses dilated convolution and FFDNet~\cite{zhang2018ffdnet} uses downsampled subimages and noise level map.


\subsection{Model based methods}
Conventional model based methods do not need training but rely on an inductive bias given as a prior. 
The performance of model based methods depends on the chosen prior knowledge.
There are several image priors such as total variation (TV)~\cite{TV_PRIOR}, Wavelet-domain processing~\cite{waveletsparsityprior} and BM3D~\cite{BM3D_2007_Dabov}.
Each prior assumes that the prior distribution is smoothness, low rank and self-similarity, respectively.

\vspace{-1.5em}\paragraph{Image prior by deep neural networks.}
Ulyanov~\etal~\cite{DIP_2020_IJCV} show that a randomly initialized convolutional neural network serves as an image prior and name it as deep image prior (DIP) and apply to several inverse problems. 

Besides the broad usages, the performance of denoining with DIP is still disappointing because of ``overfitting'' to noise (see Sec.~\ref{sec:Prelim_DIP}).
There are several remedies for the noise overfitting of DIP~\cite{DIP_2018_CVPR, Deep_heckel_2018_ICLR, GPDIP_Cheng_2019_CVPR,DIPRED_Mataev_2019_ICCV}.
DIP-RED~\cite{DIPRED_Mataev_2019_ICCV} combines a plug-and-play prior with DIP, which changes the converge point of DIP.
GP-DIP~\cite{GPDIP_Cheng_2019_CVPR} shows that DIP is asymptotically equivalent to a stationary Gaussian Process prior and introduces stochastic gradient Langevin dynamics (SGLD)~\cite{SGLD_19}.
Deep decoder~\cite{Deep_heckel_2018_ICLR} utilizes under-parameterized network based on the fact that overfitting is related to model complexity.
Inspired by that, we systematically analyze the fitting of a network but improve the performance of DIP without sacrificing the network size.

Recently, Self2Self~(S2S)~\cite{S2S_Quan_2020_CVPR} introduces self-supervised learning based on dropout and ensembling.
Owing to model uncertainty from dropout, S2S generates multiple independent denoised instance and averages the outputs for low-variance solution.
It outperforms existing solutions but needs extensive iteration with very low learning rate due to dropout.
In addition, there is an approach to combine SURE~\cite{stein_origin} with DIP~\cite{DIPSURE_metzler2020unsupervised} (DIP-SURE).
They share similarity to our work for both use SURE but we further extend it to propose a `stochastic temporal ensembling,' which deviates from the original SURE formulation.
Please find further discussion in Sec.~\ref{sec:Noise_reg}. 

\subsection{Effective degrees of freedom}
Effective degrees of freedom (DF)~\cite{df2004Efron, Tibshirani2014DoF} provides a quantitative analysis of the amount of fitting of a model to the training data.
Efron shows that an estimate of optimism is difference of error on test and training data and relates it to a measure of model complexity deemed effective degrees of freedom~\cite{df2004Efron}.
Intuitively, it reflects the effective number of parameters used by a model in producing the fitted output~\cite{Tibshirani2014DoF}.
We use the notion of DF to analyze and detect the overfitting of a network and propose our method.

\subsection{Stein's unbiased risk estimator (SURE)}
Stein's unbiased risk estimator~\cite{stein_origin} is a risk estimator for a Gaussian random variable.
It is a useful tool for selecting a model or hyper-parameters in denoising problem, since it guarantees unbiasedness for risk estimator without a target vector~\cite{Donoho95adaptingto, Zhang98AdaptiveSURErisk}.
The analytic solution for SURE is only available for limited conditions; non-local mean or linear filter~\cite{Ville2009Nonlocal, NonlocalSURE2011Ville}.
When the closed form solution is not available, Ramani~\etal~\cite{MCSURE_Ramani_2008_TIP} proposed a Monte Carlo-based SURE (MC-SURE) method to determine near-optimal parameters based on the brute-force search of the parameter space.
As the SURE based method is limited to Gaussian noise~\cite{SURE_2018_NIPS}, several works extend it to other types of noises including Poisson~\cite{PURE2011Luisier}, Poisson-Gaussian~\cite{PGURE2014Montagner}, exponential family~\cite{GSURE2009Eldar} or non-parametric noise model~\cite{BaysianSupervision2007Sch}.
We also modify our objective to extend our method to Poisson noise by~\cite{PGURE2014Montagner, PURE2011Luisier, PURE2018Soltanayev} (Sec.~\ref{sec:poisson}).


\section{Preliminaries}
\label{sec:edof}


\paragraph{Deep image prior (DIP).}
\label{sec:Prelim_DIP}

Let a noisy image $\mathbf y \in \mathbb{R}^N$ be modeled as 
\vspace{-0.5em}
\begin{equation}
    \mathbf y = \bf x + \bf n,
    \label{eq:noisemodel}
    \vspace{-0.5em}
\end{equation}
where $\mathbf x \in \mathbb{R}^N$ be a noiseless image that one would like to recover and $\mathbf n \in \mathbb{R}^N$ be an $i.i.d.$ Gaussian noise such that $\mathbf{n} \sim \mathcal{N}(\mathbf{0}, \sigma^2 \mathbf{I})$ where $\bf{I}$ is an identity matrix.
Denoising can be formulated as a problem of predicting the unknown $\mathbf{x}$ from known noisy observation $\mathbf{y}$.
Ulyanov~\etal~\cite{DIP_2020_IJCV} argued that a network architecture naturally encourages to restore the original image from a degraded image $\mathbf{y}$ and name it as deep image prior (DIP).
Specifically, DIP optimizes a convolutional neural network $\mathbf{h}$ with parameter $\boldsymbol{\theta}$ by a simple least square loss $\mathcal{L}$ as:
\begin{equation}
    \hat{\boldsymbol{\theta}} = \argmin_{\boldsymbol{\theta}} \mathcal{L}(\mathbf{h}(\dot{\mathbf{n}}; \boldsymbol{\theta}), \mathbf{y}),
    \label{eq:dip}
\end{equation}
where $\dot{\mathbf{n}}$ is a random variable that is independent of $\mathbf{y}$. 
If $\mathbf{h}$ has enough capacity (\ie, sufficiently large number of parameters or architecture size) to fit to the noisy image $\mathbf{y}$, the output of model $\mathbf{h}(\dot{\mathbf{n}}; \hat{\boldsymbol{\theta}})$ should be equal to $\mathbf{y}$, which is \emph{not} desirable.
DIP uses the early stopping to obtained the results with best PSNR with clean images.



\vspace{-1em}\paragraph{Effective degrees of freedom for DIP.}
\label{sec:edof}
The effective degrees of freedom~\cite{df2004Efron, Tibshirani2014DoF} quantifies the amount of fitting of a model to training data.
We analyze the training of DIP by the effective degrees of freedom (DF) in Eq.~\ref{eq:df} as a tool for monitoring overfitting to the given noisy image.
the DF for the estimator $\mathbf{h}(\cdot)$ of $\mathbf{x}$ with input $\mathbf{y}$ can be defined as follows~\cite{hastie1990generalized}:
\begin{equation}
\vspace{-0.5em}
    \text{DF}(\mathbf{h})= \frac{1}{\sigma^2} \sum_{i=1}^n Cov(\mathbf{h}_i(\cdot), \mathbf{y}_i),
    \label{eq:df}
\end{equation}
where $\mathbf{h}(\cdot)$ and $\mathbf{y}$ are a model (\eg, a neural network) and noise image respectively. $\sigma$ is the standard deviation of the noise. $\mathbf{h}_i(\cdot)$ and $\mathbf{y}_i$ indicate the $i^\text{th}$ element of corresponding vectors. 
For example, if the input to $\mathbf{h}(\cdot)$ is $\dot{\mathbf{n}}$ and $\mathbf{y}$ is a noisy image, \ie, $\mathbf{h}(\dot{\mathbf{n}})$, it is the DF for DIP.
Note that $\mathbf{h}(\cdot)$ can take any input and we use $\mathbf{y}$ (instead of $\dot{\mathbf{n}}$) for our formulation.

Interestingly, the DF is closely related to the notion of \emph{optimism} of an estimator $\mathbf{h}$, which is defined by the difference between test error and train error~\cite{hastie1990generalized, Tibshirani2014DoF} as:
\begin{equation}
\vspace{-0.5em}
    \rho(\mathbf{h}) = \mathbb{E} \left[ \mathcal{L}(\tilde{\mathbf{y}}, \mathbf{h}(\cdot)) - \mathcal{L}(\mathbf{y}, \mathbf{h}(\cdot))\right],
\label{eq:optimism}
\end{equation}
where $\mathcal{L}(\cdot)$ is a mean squared error (MSE) loss, $\tilde{\mathbf{y}}$ is another realization from the model (\ie, with different $\mathbf{n}$ in Eq.~\ref{eq:noisemodel}) that is independent of $\mathbf{y}$.
In~\cite{Tibshirani2014DoF}, it is shown that $\rho(\mathbf{h}) = 2 \sum_{i=1}^n Cov(\mathbf{h}_i(\cdot), \mathbf{y}_i)$. 
Thus, combining with Eq.~\ref{eq:df}, it is straightforward to show that
\begin{equation}
\vspace{-0.5em}
    2 \sigma^2 \cdot \text{DF}(\mathbf{h}) = \rho(\mathbf{h}).
\end{equation}

It is challenging to compute the covariance since $\mathbf{h}(\cdot)$ is nonlinear (\eg, a neural network), gradually changing in optimization, and the $\rho(\mathbf{h})$ requires many pairs of noisy and clean (ground-truth) images to compute (note that it is an estimate).
Here, we introduce a simple approximated degrees of freedom with a single ground-truth and call it as $\text{DF}_{GT}$.
We derive the $\text{DF}_{GT}$ as following:
\begin{equation}
\begin{split}
2\sigma^2 \cdot \text{DF}_{GT}(\mathbf{h}) 
    \approx \mathcal{L}(\mathbf{x}, \mathbf{h(\cdot)}) - 
    \mathcal{L}(\mathbf{y}, \mathbf{h(\cdot)}) + \sigma^2
\vspace{-0.5em}
\label{eq:dfgt}
\end{split}
\end{equation}
We describe a simple proof of the estimation in the supplementary material.

A large DF implies overfitting to the given input $\mathbf{y}$, which is not desirable.
If DIP fits to $\mathbf{x}$, $\text{DF}_{GT}$ becomes close to 0.
The more the DIP is fitting to $\mathbf y$, the larger the DF is.
We use the $\text{DF}_{GT}$ to analyze the DIP optimization in empirical studies in Sec.~\ref{sec:conv_df_gt}.

\section{Approach}
\label{sec:Approach}
To prevent the overfitting of DIP, we try to suppress the DF~(Eq.~\ref{eq:df}) during the optimization without the access of ground-truth clean image $\mathbf{x}$.
In Eq.~\ref{eq:df}, computing the DF is equivalent to the sum of the covariances for each element of the noise image $\mathbf{y}$ and the model output $\mathbf{h}(\cdot)$.
There are a number of techniques to simply approximate the covariance computation in statistical learning literature such as AIC~\cite{AIC1973akaike}, BIC~\cite{BIC1978Schwarz} and Stein's unbiased risk estimator (SURE)~\cite{stein_origin}.
Both AIC and BIC, however, approximate the DF by counting the number of parameters of a model, so for usual over-parameterized deep neural networks, the approximations based on them could be incorrect~\cite{modelselectionNN1999Ulrich}.
Note that $\text{DF}_{GT}$ cannot be used for optimizing model because it needs groud-truth clean image $\mathbf{x}$.

Here, we propose to use SURE to suppress the DF by deriving the DIP formulation using the Stein's lemma. 
The Stein's lemma for a multivariate Gaussian vector $\mathbf{y}$ is~\cite{stein_origin}:
\begin{equation}
\frac{1}{\sigma^2} \sum_{i=1}^n Cov(\mathbf{h}_i(\mathbf{y}), \mathbf{y}_i) = 
\mathbb E \left[ \sum_{i=1}^n \frac{\partial \mathbf{h}_i(\mathbf{y})}{\partial \mathbf{y}_i} \right].
\label{eq:steinlemma}
\end{equation}
It simplifies the computation of DF from the covariances between $\mathbf{y}$ and $\mathbf{h}(\mathbf{y})$ to the expected partial derivatives at each point, which is well approximated in a number of computationally efficient ways~\cite{MCSURE_Ramani_2008_TIP, NEWDIV_Soltanayev_2020_ICASSP}. 
Note that the SURE which is denoted as $\eta(\mathbf{h}(\mathbf{y}),\mathbf{y})$, consists of Eq.~\ref{eq:steinlemma} and the DIP loss (Eq.~\ref{eq:dip}) with a modification of its input (from $\dot{\mathbf{n}}$ to $\mathbf{y}$) as:
\begin{equation}
\vspace{-0.5em}
\resizebox{0.9\linewidth}{!}{
   $\eta(\mathbf{h}(\mathbf{y}),\mathbf{y})$ =
    $\mathcal{L}(\mathbf{y}, \mathbf{h}(\mathbf{y})) + \underbrace{\frac{2\sigma^2}{N}\sum^N_{i=1}\frac{\partial \mathbf{h}_i(\mathbf{y})}{\partial (\mathbf{y})_i}}_{\text{divergence term}} - \sigma^2$.
    }
\label{eq:sure_def}
\end{equation}
While the vanilla DIP loss encourages to fit the output of the model $\mathbf{h}$ to noisy image $\mathbf{y}$, Eq.~(\ref{eq:sure_def}) encourages to approximately fit it to clean image $\mathbf{x}$ without access to the $\mathbf{x}$.

However, it is still computationally demanding to use Eq.~\ref{eq:sure_def} as a loss for optimization with any gradient based algorithm due to the divergence term~\cite{MCSURE_Ramani_2008_TIP}.
A Monte-Carlo approximation for Eq.~\ref{eq:sure_def} in~\cite{MCSURE_Ramani_2008_TIP} can be a remedy to the computation cost, but it introduces a hyper-parameter $\epsilon$ that has to be selected properly for the best performance on different network architectures and/or datasets. 
For not requiring to tune the hyper-parameter $\epsilon$, we employed an alternative Monte-Carlo approximation for the divergence term~\cite{NEWDIV_Soltanayev_2020_ICASSP} as:
\begin{align}
    \frac{1}{N}\sum^N_{i=1}\frac{\partial \mathbf{h}_i(\mathbf{y})}{\partial \mathbf{y}_i} \approx \frac{1}{N}\tilde{\mathbf{n}}^T \mathbf{J}_{\tilde{\mathbf{n}}^T \mathbf{h(y)}},
    \label{eq:MCdiv2}
\end{align}
where $\tilde{\mathbf{n}}$ is a standard normal random vector, \ie, $\tilde{\mathbf{n}} \sim \mathcal{N}(\mathbf{0}, \mathbf{I})$ and the $i^\text{th}$ element of the Jacobian $\mathbf{J}_{\tilde{\mathbf{n}}^T \mathbf{h(y)}}$ 
is $\partial \tilde{\mathbf{n}}^T\mathbf{h(y;\theta)/\partial y}_i$.
We denote this `estimated degrees of freedom by Monte-Carlo' by $\text{DF}_{MC}$ and will use it to monitor the DIP optimization without using the PSNR with the clean ground truth images (Sec.~\ref{sec:stopping_criterion}).


\subsection{Stochastic temporal ensembling}
To improve the fitting accuracy, DIP suggests several methods including noise regularization, exponential moving average~\cite{DIP_2020_IJCV}.
We propose `stochastic temporal ensembling (STE)' for better fitting performance by leveraging these methods to our objective.

\vspace{-1em}\paragraph{Noise regularization on DIP.}
\label{sec:Noise_reg}
DIP shows that adding extra temporal noises to the input $\dot{\mathbf{n}}$ of function $\mathbf{h}(\cdot)$ at each iteration improves performance for the inverse problems including image denoising~\cite{DIP_2020_IJCV}.
It is to add a noise vector $\gamma$, with $\gamma \sim N(0, \sigma^2_\gamma \mathbf I)$ to the input of the function at every iteration of the optimization as:
\begin{equation}
    \hat{\boldsymbol{\theta}} = \argmin_{\boldsymbol{\theta}} \mathcal{L}(\mathbf{h}(\dot{\mathbf{n}} + \gamma; \boldsymbol{\theta}), \mathbf{y}),
    \label{eq:noise_reg}
\end{equation}
where $\dot{\mathbf{n}}$ is fixed but $\gamma$ is sampled from Gaussian distribution with zero mean, standard deviation of $\sigma_{\gamma}$ at every iteration.
%
To estimate the $\mathbf{x}$ by Eq.~\ref{eq:sure_def}, we replace the input of the model $\mathbf{h}(\cdot)$, $\dot{\mathbf{n}}$, with noisy image $\mathbf{y}$ (from Eq.~\ref{eq:df} to Eq.~\ref{eq:steinlemma}).
Interestingly, Eq.~\ref{eq:noise_reg} becomes similar to the denoising auto-encoder (DAE), which prevents a model from learning a trivial solution by perturbing output of $\mathbf{h}$~\cite{DAE2008Vincent}.

Meanwhile, contractive autoencoder (CAE)~\cite{CAE2011Rifai} minimizes the Frobenius norm of the Jacobian and SURE and its variants minimize the trace of the Jacobian (Eq.~\ref{eq:MCdiv2}) thus suppresses the DF.
Since we assume that the different realizations of noise are independent, the off-diagonal elements of the matrix are zero, CAE is equivalent to SURE in terms of suppressing the DF. 
Alain~\etal~\cite{regautoencoder2014alain} later show that the DAE is a special case of the CAE when $\sigma_{\gamma} \rightarrow 0$.
We can rewrite the Eq.~\ref{eq:noise_reg} by using CAE formulation as:
\begin{equation}
    \argmin_{\boldsymbol{\theta}} \mathcal{L}(\mathbf{h}(\mathbf{y}; \boldsymbol{\theta}), \mathbf{y}) + \sigma_{\gamma}^2 \Bigg\| \frac{\partial \mathbf{h}(\mathbf{y}; \boldsymbol{\theta})}{\partial \mathbf{y}} \Bigg\|_F^2 + o(\sigma^2_\gamma),
    \label{eq:noise_reg_derivative}
\end{equation}
when~$\sigma_\gamma \rightarrow 0$, where $o(\sigma^2_\gamma)$ is a high order error term from Taylor expansion.
Thus, solving this optimization problem is equivalent to penalizing increase of DF.
Here, the noise level $\sigma_\gamma$ serves as a hyper-parameter for determining performance and it improves performance of DIP by using multiple level of $\sigma_z$ at optimization of DIP.
Thus, we further proposed to model $\sigma_\gamma$ as a uniform random variable instead of a empirically chosen hyper-parameter such that
\begin{equation}
\vspace{-0.5em}
    \sigma_\gamma \sim \mathcal U(0, b).
    \label{eq:ste}
    \vspace{-0.5em}
\end{equation}

\vspace{-1em}\paragraph{Exponential moving average.}
DIP further shows that averaging the restored images obtained in the last iterations improves the performance of denoising~\cite{DIP_2020_IJCV}, which we refer to as `exponential moving average (EMA).'
It can be thought as an analogy to the effect of ensembling~\cite{datadistill2018}.

\vspace{-1em}\paragraph{Stochastic temporal ensembling.}
Leveraging the noise regularization and the EMA, we propose a method called `stochastic temporal ensembling (STE)' to improve the fitting performance of DIP loss.
%
Specifically, we modify our formulation (Eq.~\ref{eq:sure_def}) by allowing two noise observations, $\mathbf{y_1}$ for target of MSE loss and $\mathbf{y_2}$ for the input of the model, $h$, instead of one $\mathbf{y}$ by setting $\mathbf{y}_1 = \mathbf{y}$ and $\mathbf{y}_2= \mathbf{y}+\gamma$ as:
\begin{equation}
\resizebox{0.88\linewidth}{!}{
$   \eta(\mathbf{h}(\mathbf{y_2}),\mathbf{y_1}) =
    \underbrace{\mathcal{L}(\mathbf{h}(\mathbf{y_2}), \mathbf{y_1})}_{\text{data fidelity}} + \underbrace{\frac{2\sigma^2}{N}\sum^N_{i=1}\frac{\partial \mathbf{h}_i(\mathbf{y_2})}{\partial (\mathbf{y_2})_i}}_{\text{regularization}} - \sigma^2,$
    }
\label{eq:esure_def}
\end{equation}
where $\sigma$ is a known noise level of $\mathbf{y_1}$ (same as Eq.~\ref{eq:noisemodel}), $\mathbf{h}_i(\mathbf{y_2})$ and $(\mathbf{y_2})_i$ are the $i^\text{th}$ element of the vectors of $\mathbf{h}(\mathbf{y_2})$ and $\mathbf{y_1}$, respectively.
%
Interestingly, Eq.~\ref{eq:esure_def} is equivalent to the formulation of extended SURE (eSURE)~\cite{eSURE2019Neurips}, which is shown to be a better unbiased estimator of the MSE with the clean image $\mathbf x$.
But there are a number of critical differences of ours from ~\cite{eSURE2019Neurips}.
First, Our method does not require training, while Zhussip~\etal~\cite{eSURE2019Neurips} requires training with many noisy images.
Because Zhussip~\etal~\cite{eSURE2019Neurips} use the fixed instance of $\gamma$, there is no effect of regularization from (Eq.~\ref{eq:noise_reg}), which gives reasonable performance gain( See Sec.\ref{sec:exp_quanti}).
This is our final objective function of DIP that stops automatically by a stopping criterion, described in the following section.

\subsection{Zero-crossing stopping criterion}
\label{sec:stopping_criterion}

SURE works well if the model $\mathbf{h}$ satisfies the smoothness condition, \ie, $\mathbf{h}$ admits a well-defined second-order Taylor expansion~\cite{SURE_2018_NIPS,DIPSURE_metzler2020unsupervised}.
While a typical learning based denoiser satisfies this smoothness condition~\cite{SURE_2018_NIPS,eSURE2019Neurips}, the DIP network `fits' to a target image (a noisy image in~\cite{DIP_2018_CVPR,DIP_2020_IJCV} and an approximate clean image in our objective) and therefore there is no guarantee that the smoothness condition can be satisfied, especially when it has been converged.

We observed that the divergence term in our formulation (Eq.~\ref{eq:esure_def}) increases at early iterations (\ie, before convergence) while it starts to diverge to $-\infty$ at later iterations (\ie, after convergence).
This observation is consistent in all our experiments. 
Note that this divergence phenomenon was not reported in~\cite{DIPSURE_metzler2020unsupervised} because the DIP network with the SURE loss did not seem to be fully converged to recover the fine details with insufficient number of iterations.
Based on this observation for our proposed objective, we propose `zero crossing stopping criterion' to stop iteration when our objective function (Eq.~\ref{eq:esure_def}) deviates from zero.


\vspace{-1em}\paragraph{Solution trajectory.}
\label{sec:sol_traj}

\begin{figure}[t!]
    \centering
    \includegraphics[width=1.0\linewidth]{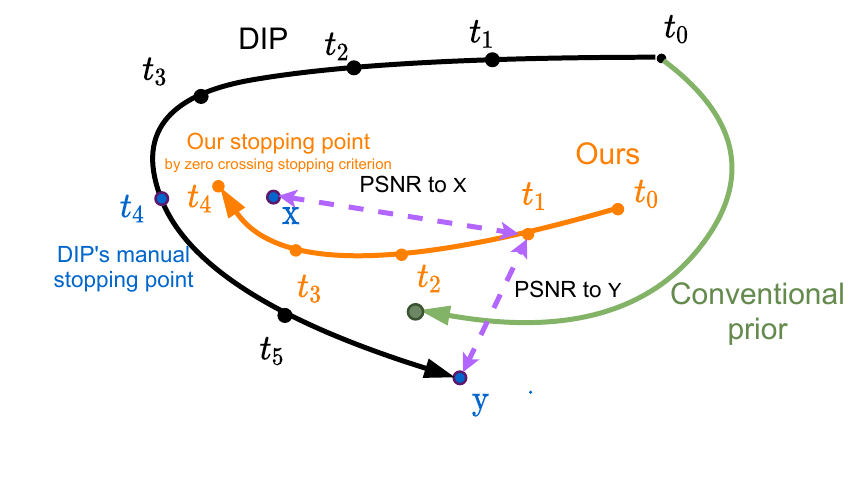}
    \vspace{-3.0em}
    \caption{\textbf{Illustration of a solution trajectory of ours and DIP.} 
    We consider the problem of reconstructing an image $\mathbf{x}$ from a degraded measurement $\mathbf{y}$.
    DIP finds its optimal stopping point ($t_4$) by early stopping. Ours changes DIP's solution trajectory from black to orange whose stopping point ($t_5$) is defined by a loss value (Sec.\ref{sec:stopping_criterion}) and is close to noiseless solution ($\textbf{x}$).
    }
    \vspace{-1em}
    \label{fig:solution}
\end{figure}
\label{sec:background}

To help understand the difference between our method to DIP in optimization procedure, similar to Fig.~3 in~\cite{DIP_2020_IJCV}, we illustrate DIP image restoration trajectory with that of our method in Fig.~\ref{fig:solution}.
DIP degrades the quality of the restored images by the overfitting.
To obtain the solution close to the clean ground truth image, DIP uses early stopping ({\color{blue}blue $t_4$}).
Our formulation has different training trajectory ({\color{orange}orange}) from DIP (black) and automatically stops the optimization by the zero crossing stopping ({\color{orange} orange $t_4$}).
We argue that the resulting image by our formulation is in general closer to the clean image ({\color{blue}blue $\mathbf{x}$}) than the solution by DIP, which preserves more high frequency details than the solution by the DIP (Sec.~\ref{sec:exp_quali}) thanks to a better target to fit (an approximation of the clean $\mathbf{x}$ over a noisy image $\mathbf{y}$ and our proposed principled stopping criterion without using ground truth image).
We empirically analyze this phenomenon with our proposed $DF_{GT}$ and compare it to $DF_{MC}$ in Sec.~\ref{sec:conv_df_gt} and the supplementary material.

\subsection{Extension to Poisson noise}
\label{sec:poisson}
As the SURE is limited to Gaussian noise~\cite{SURE_2018_NIPS}, there are several attempts to extend it to other types of noises~\cite{PGURE2014Montagner,GSURE2009Eldar,BaysianSupervision2007Sch}.
Here, we extend our formulation to Poisson noise as it is a useful model for noise in low-light condition.
We modify our formulation (Eq.~\ref{eq:esure_def}) to use Poisson unbiased risk estimator (PURE)~\cite{PURE2011Luisier,PGURE2014Montagner,Kim2020PURECT} for Poisson noise as follows:
\begin{equation}
\resizebox{0.88\linewidth}{!}{
    $\mathcal{L}(\mathbf{h}(\mathbf{y}), \mathbf{y}) - \frac{\zeta}{N} \sum^N_{i=1} \mathbf{y}_i $\\
    $+\frac{2\zeta}{\dot{\epsilon}N}(\mathbf{\Breve{n} \odot y)^T(h(y+\dot{\epsilon}\Breve{\mathbf{n}}) - \mathbf{h(y)}))}$,
}
\label{eq:poisson}
\end{equation}
where $\mathbf{\Breve{n}}$ is a $k$-dimensional binary random variable whose element $\Breve{n}_i$ takes -1 or 1 with probability 0.5 for each, $\dot{\epsilon}$ is a small positive number and $\odot$ is a Hadamard product. 
We empirically validate the Poisson extension in Sec.~\ref{sec:poisson_exp}.

\section{Experiments}

\begin{figure*}[t!]
\centering
    \begin{subfigure}{.33\textwidth}
    \centering
    \vspace{-1.2em}
    \includegraphics[width=1\textwidth]{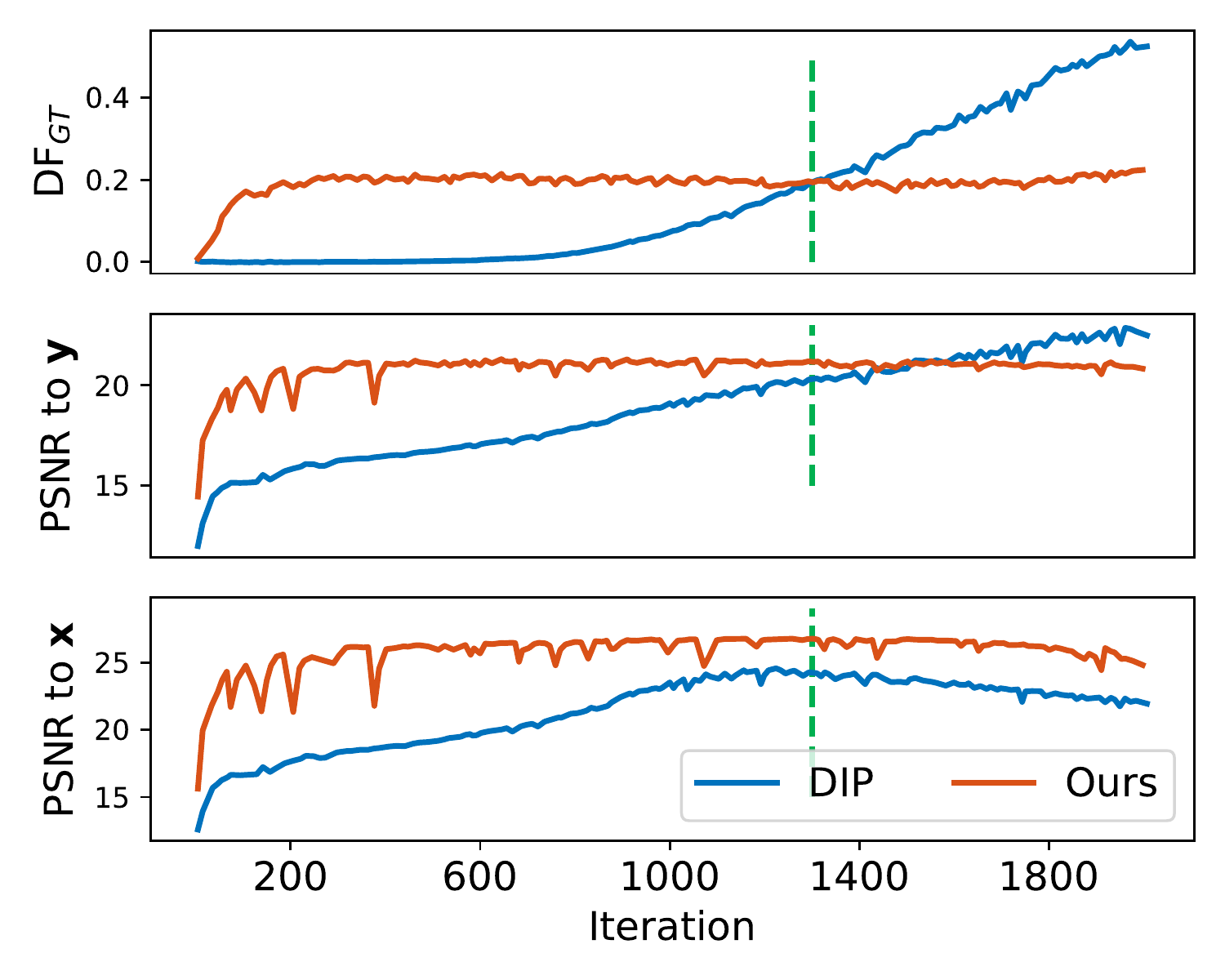}
    \vspace{-2em}
    \caption{Convergence analysis}
    \label{fig:df_dip_ours}
\end{subfigure}
\begin{subfigure}{.33\textwidth}
    \centering
    \vspace{-1.2em}
    \includegraphics[width=1\textwidth]{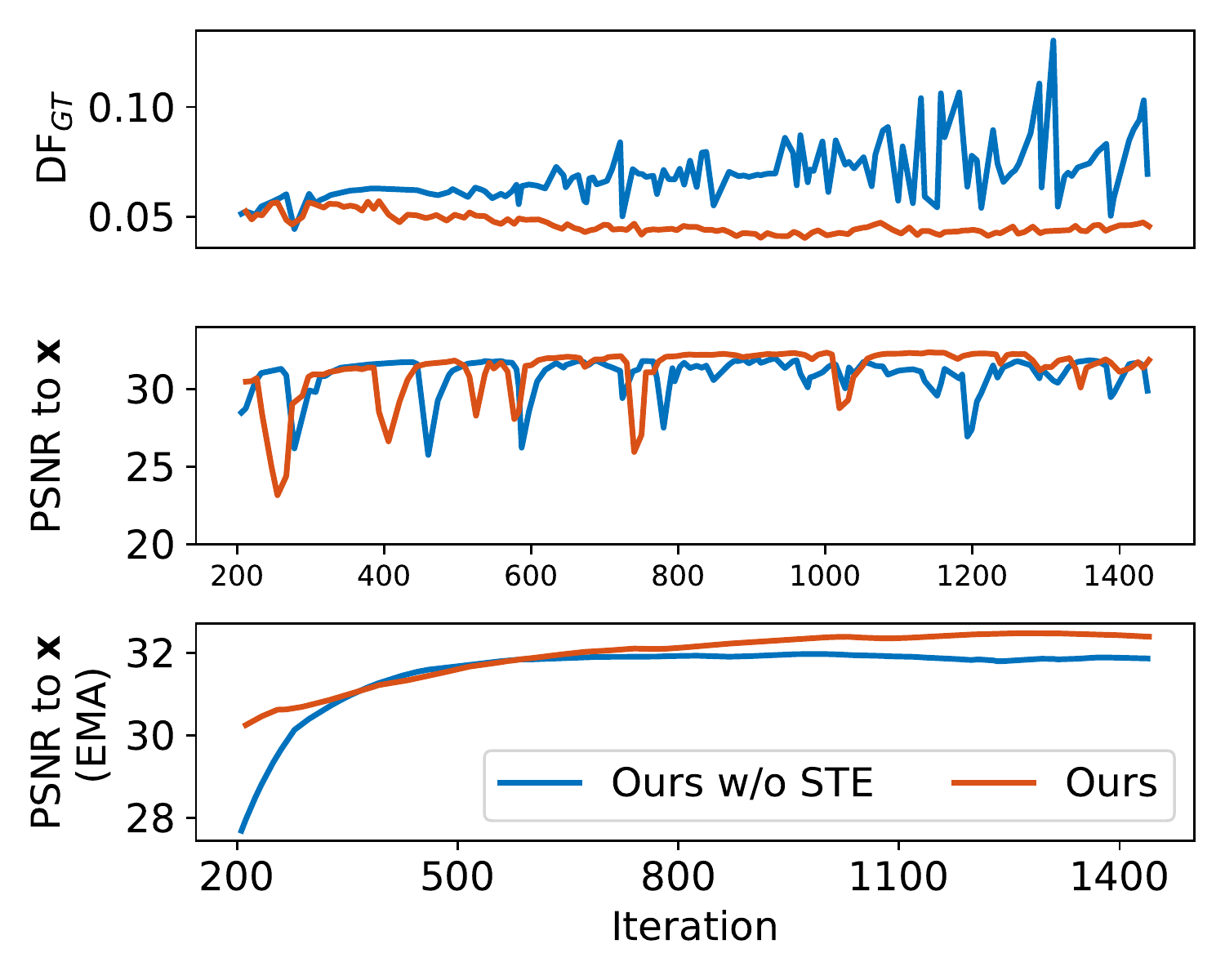}
    \vspace{-2em}    
    \caption{Effect of the STE}
    \label{fig:sure_esure}
\end{subfigure}
\begin{subfigure}{.33\textwidth}
    \centering
    \vspace{-1.2em}
    \includegraphics[width=1\textwidth]{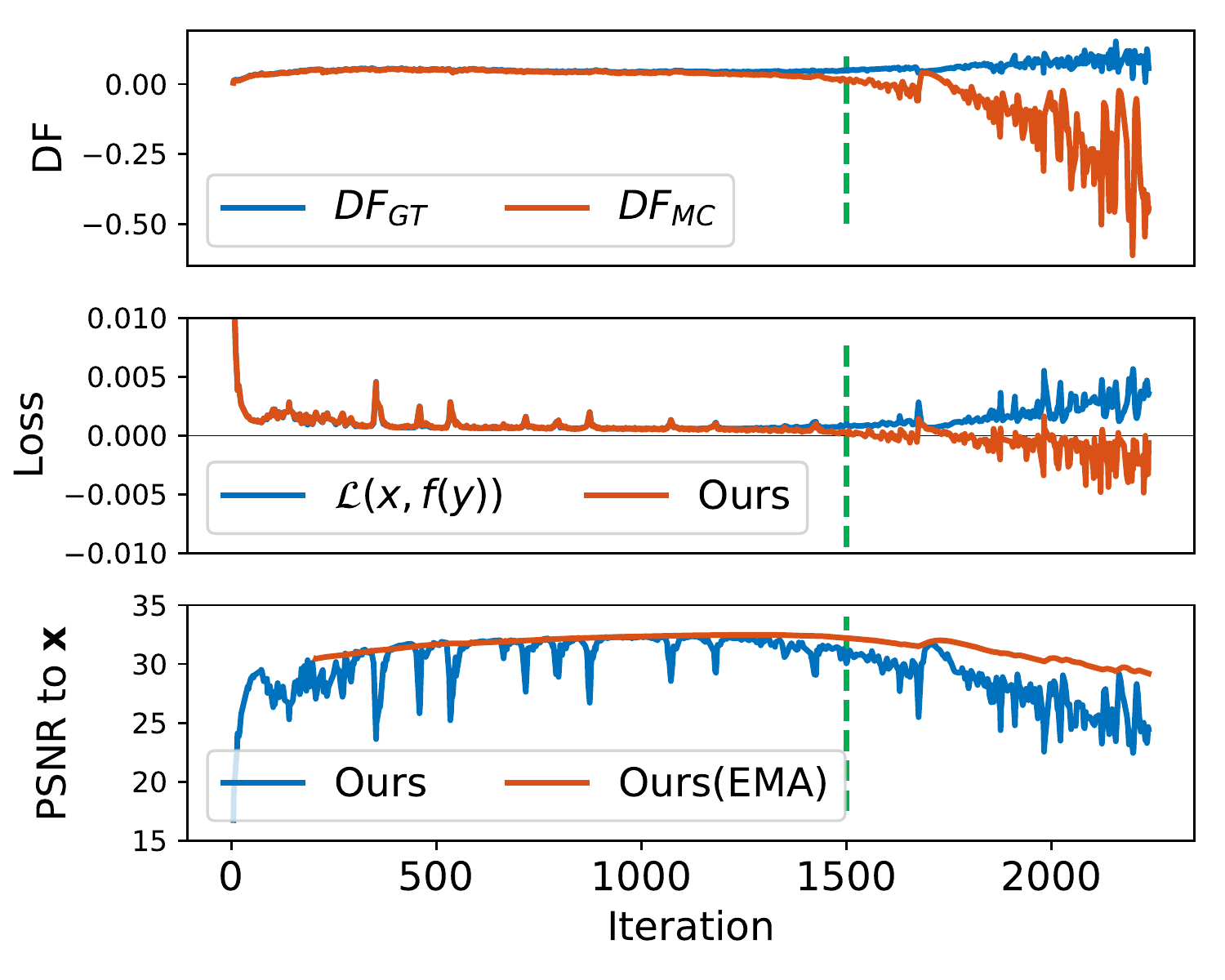}
    \vspace{-2em}    
    \caption{Zero-crossing stopping criterion}
    \label{fig:SURE_OURS}
\end{subfigure}
\caption{\textbf{Learning analysis with $\text{DF}_{GT}$}. 
    (a) As optimization progresses, degrees of freedom of DIP increase as fit to noisy observation. Ours does not overfit to noisy observation and shows consistently better performance thanks to Stein unbiased risk estimation. The {\color{darkgreen}green dashed line} indicates the intersection between DIP and ours in $\text{DF}_{GT}$.
    (b) The proposed method makes optimization more stable than single instance one and this tendency also is observed on PSNR to $\mathbf{x}$, PSNR to $\mathbf{x}$(EMA).
    (c) Monte-carlo estimation $\text{DF}_{MC}$  of eSURE is stable for a considerable amount of steps but the error of estimation is soaring. It usually happens  when the loss is already close to zero (the {\color{darkgreen}green dashed line} on the plot). Thus, we propose to stop the optimization as soon as the loss reaches zero.
    }
    \vspace{-1em}
    \label{fig:stopping}
\end{figure*}


\paragraph{Implementation details.} 
\label{sec:Implementation_details}
For the $\sigma$, we use $\sigma=15, 25, 50$, following~\cite{DnCNN_2017_TIP}, and $\sigma=25$ for in-depth analysis.
For the $b$ in Eq.~\ref{eq:ste}, we set it to the same value to the $\sigma$.
RAdam opimizer~\cite{liu2019radam} is used for training with learning rate 0.1.
Details including network architectures and datasets are in the supplementary material.

\vspace{-1em}\paragraph{Evaluation metrics.}
We use peak signal-to-noise ratio (PSNR), structured similarity (SSIM) and learned perceptual image patch similarity (LPIPS)~\cite{LPIPS_zhang2018perceptual}.
The PSNR is widely used in denoising literature~\cite{DnCNN_2017_TIP, zhang2018ffdnet,IRCNNzhang_2017_learning, N3Net_Pl_2018, jia2019focnet} but is recently argued that it is not an ideal metric as it values the oversmoothed results~\cite{LPIPS_zhang2018perceptual, Ledig2017SRGAN}.
For this reason, we compare the algorithms with LPIPS as an alternative measurement of human study.
We use the publicly available pre-trained weights based on AlexNet by the authors~\cite{LPIPS_zhang2018perceptual}.
We additionally report the performance of the peak PSNR during optimization of our method as a reference (denoted as `Ours*').

\subsection{Convergence analysis by $\text{DF}_{GT}$}
\label{sec:conv_df_gt}
Fig.~\ref{fig:df_dip_ours} shows the $\text{DF}_{GT}$, PSNR to $y$ and PSNR to $x$; $\text{DF}_{GT}$ is the effective degrees of freedom with Ground Trues, PSNR to $y$ and PSNR to $x$ refer PSNR from the model output to $y$ and $x$ respectively.
As optimization progresses, the degrees of freedom of DIP increases gradually with PSNR to $\mathbf{x}$.
But PSNR to $\mathbf{x}$ of DIP decreases from 1,300 iteration.
In contrast, in ours, $\text{DF}_{GT}$ rises at the beginning of the iterations and stays at a certain value.
%
Interestingly, the best stopping point for DIP is near the intersection between DIP and our method in $\text{DF}_{GT}$.
It implies that the converged $\text{DF}_{GT}$ value by our method is near the optimal solution of DIP ($t_4$ of DIP in Fig.~\ref{fig:solution}).

Fig.~\ref{fig:sure_esure} shows the trajectory of two objectives on $\text{DF}_{GT}$; (1) ours w/o STE and (2) ours.
As shown in Sec~\ref{sec:Noise_reg}, STE suppresses the DF by minimizing the norm of the Jacobian, which is similar to trace of Jacobian (the DF from the Stein's lemma).
Accordingly, ours suppresses the DF better than ours w/o STE in $\text{DF}_{GT}$ (Fig.~\ref{fig:sure_esure} (top)).
This tendency is also observed in `PSNR to $\mathbf{x}$' and `PSNR to $\mathbf{x}$ (EMA).'

The optimization progresses of $DF_{GT}$ and $DF_{MC}$ are shown in Fig.~\ref{fig:SURE_OURS}.
The $DF_{MC}$ starts underestimating the $DF_{GT}$ after a certain iteration and Eq.~ \ref{eq:esure_def} (`Loss') becomes zero and it reaches the highest PSNR value.
Thus, our proposed zero stopping criterion detects when $DF_{MC}$ fails to estimate the $DF_{GT}$; when the loss crosses zero.

\subsection{Quantitative analysis}
\label{sec:exp_quanti}
\begin{table}[t]
\vspace{-0.5em}
\resizebox{1.0\linewidth}{!}{
\begin{tabular}{ccccc}
\toprule
\textbf{Method} & \textbf{Overfit Prev.} & \textbf{PSNR ($\uparrow$)}  & \textbf{SSIM ($\uparrow$)} & \textbf{LPIPS ($\downarrow$)} \\ \midrule
DIP~\cite{DIP_2020_IJCV}       & Early stopping           & 29.96 & 0.940 & 0.152 \\
Deep Decoder~\cite{Deep_heckel_2018_ICLR}        & Under-param.       & 26.94                         & 0.889                         & 0.377 \\
DIP-RED~\cite{DIPRED_Mataev_2019_ICCV}       & Plug-and-play            & 30.88                         & 0.932                         & 0.197 \\
GP-DIP~\cite{GPDIP_Cheng_2019_CVPR}        & SGLD                     & 29.99                         & 0.948                         & 0.251 \\ 
DIP-SURE*~\cite{DIPSURE_metzler2020unsupervised}        & ZCSC                     & 30.33                         & 0.941                         & 0.149 \\\midrule
\rowcolor[HTML]{EFEFEF} 
{\bf Ours} w/o STE~\cite{eSURE2019Neurips}  & ZCSC       & \underline{31.34}                         & \textbf{0.955}                         & \underline{0.108} \\
\rowcolor[HTML]{EFEFEF} 
{\bf Ours}      & ZCSC      & \textbf{31.54}                         &\underline{0.953}                         & \textbf{0.107}\\
\bottomrule
\end{tabular}
}
\caption{\textbf{Comparison to DIP variants on CSet9 dataset ($\sigma=25$).} $(\uparrow)$: higher the better, $(\downarrow)$: lower the better. `Overfit Prev.' refers to `overfitting prevention method.' `Under-param.' refers to `under parameterized.' (Best values: in \textbf{bold}. The second best values: \underline{underlined}). `ZCSC' refers to the proposed zero crossing stopping criterion. `DIP-SURE*' refers to~\cite{DIPSURE_metzler2020unsupervised} with ZCSC for fair comparison among the methods using the SURE formulation.}
\vspace{-1.5em}
\label{tab:dips}
\end{table}

\begin{table*}[t]
\resizebox{1.0\linewidth}{!}{
\begin{tabular}{cccccacccaccca}
\toprule
                        &       & \multicolumn{4}{c}{PSNR ($\uparrow$)}              & \multicolumn{4}{c}{SSIM ($\uparrow$)}              & \multicolumn{4}{c}{LPIPS ($\downarrow$)}             \\
\cmidrule(lr){3-6} \cmidrule(lr){7-10} \cmidrule(lr){11-14}
                        Dataset & $\sigma$ & BM3D~\cite{BM3D_2007_Dabov}  & DIP~\cite{DIP_2020_IJCV}   & S2S~\cite{S2S_Quan_2020_CVPR}   & Ours (Ours*) & BM3D~\cite{BM3D_2007_Dabov}  & DIP~\cite{DIP_2020_IJCV}   & S2S~\cite{S2S_Quan_2020_CVPR}   & Ours (Ours*) & BM3D~\cite{BM3D_2007_Dabov}  & DIP~\cite{DIP_2020_IJCV}   & S2S~\cite{S2S_Quan_2020_CVPR}   & Ours (Ours*) \\
                        \midrule
\multicolumn{14}{l}{ \bf {\fontfamily{lmss}\selectfont \hspace{-0.5em} Color Image Datasets}}\\                        
\multirow{3}{*}{CSet9}  & 15 & \underline{33.83} & 31.83 & 33.24 & \underline{33.83} (\textbf{34.07}) & 0.972 & 0.960 & 0.968 & \underline{0.973} (\textbf{0.975}) & 0.111 & 0.114 & 0.135 & \textbf{0.070} (\underline{0.077}) \\
                        & 25 & 31.68 & 29.96 & \underline{31.72} & 31.54 (\textbf{31.88}) & \underline{0.956} & 0.940 & \underline{0.956} & 0.953 (\textbf{0.960}) & 0.161 & 0.152 & 0.173 & \textbf{0.107} (\underline{0.118}) \\
                        & 50 & 28.92 & 27.42 & \textbf{29.25} & 28.90(\underline{29.03}) & 0.922 & 0.900 & \underline{0.928} & 0.923 (\textbf{0.930}) & 0.267 & 0.291 & 0.235 & \textbf{0.181} (\underline{0.200}) \\
                        \cmidrule(lr){1-2} \cmidrule(lr){3-6} \cmidrule(lr){7-10} \cmidrule(lr){11-14}
\multirow{3}{*}{CBSD68} & 15 & \underline{33.51} & 31.48 & 32.78 & 33.43 (\textbf{33.56}) & \underline{0.961} & 0.941 & 0.956 & \underline{0.961} (\textbf{0.963}) & 0.081 & 0.081 & 0.102 & \underline{0.060} (\textbf{0.057}) \\
                        & 25 & \underline{30.70} & 28.66 & 30.67 & 30.67 (\textbf{30.86}) & \underline{0.932} & 0.900 & \underline{0.932} & \underline{0.932} (\textbf{0.936}) & 0.148 & 0.156 & 0.147 & \underline{0.102} (\textbf{0.100}) \\
                        & 50 & 27.37 & 25.70 & \textbf{27.62} & 27.43 (\underline{27.58}) & 0.871 & 0.832 & \underline{0.879} & 0.873 (\textbf{0.881}) & 0.298 & 0.329 & 0.244 & \textbf{0.194} (\underline{0.198}) \\
                        \cmidrule(lr){1-2} \cmidrule(lr){3-6} \cmidrule(lr){7-10} \cmidrule(lr){11-14}
\multirow{3}{*}{Kodak}  & 15 & \underline{34.41} & 32.17 & 33.70 & 34.35 (\textbf{34.49}) & \underline{0.962} & 0.941 & 0.958 & 0.961 (\textbf{0.963}) & 0.104 & 0.105 & 0.118 & \underline{0.080} (\textbf{0.077}) \\
                        & 25 & \underline{31.82} & 29.68 & 31.79 & 31.60 (\textbf{31.98}) & 0.938 & 0.907 & \underline{0.939} & 0.932 (\textbf{0.941}) & 0.161 & 0.173 & 0.159 & \textbf{0.117} (\underline{0.118}) \\
                        & 50 & 28.62 & 26.77 & \textbf{29.08} & 28.58 (\underline{28.76}) & 0.886 & 0.843 & \underline{0.898} & 0.882 (\textbf{0.892}) & 0.287 & 0.338 & 0.235 & \textbf{0.203} (\underline{0.209}) \\
                        \cmidrule(lr){1-2} \cmidrule(lr){3-6} \cmidrule(lr){7-10} \cmidrule(lr){11-14}
\multirow{3}{*}{McM}    & 15 & 34.05 & 32.54 & 33.92 & \underline{34.13} (\textbf{34.35}) & \underline{0.969} & 0.956 & 0.968 & 0.967 (\textbf{0.970}) & 0.068 & 0.067 & 0.089 & \underline{0.053} (\textbf{0.052}) \\
                        & 25 & 31.66 & 30.09 & \textbf{32.15} & 31.89 (\underline{31.98}) & 0.950 & 0.929 & \textbf{0.955} & 0.950 (\underline{0.953}) & \underline{0.107} & 0.123 & 0.117 & \textbf{0.085}  (\textbf{0.085}) \\
                        & 50 & 28.51 & 27.06 & \textbf{29.29} & \underline{28.83} (28.82) & 0.910 & 0.882 & \textbf{0.924} & 0.913 (\underline{0.918}) & 0.207 & 0.252 & 0.178 & \textbf{0.151} (\underline{0.162}) \\
                        \midrule
\multicolumn{14}{l}{ \bf {\fontfamily{lmss}\selectfont \hspace{-0.5em} Gray-scale Image Datasets}}\\
\multirow{3}{*}{BSD68}  & 15 & \underline{31.07} & 28.83 & 30.62 & 30.98 (\textbf{31.21}) & 0.872 & 0.812 & 0.858 & \underline{0.873} (\textbf{0.882}) & 0.147 & 0.163 & 0.163 & \textbf{0.090} (\underline{0.099}) \\
                        & 25 & 28.57 & 26.59 & \underline{28.60} & 28.40 (\textbf{28.78}) & \underline{0.801} & 0.734 & \underline{0.801} & 0.800 (\textbf{0.818}) & 0.226 & 0.262 & 0.197 & \textbf{0.157} (\underline{0.159}) \\
                        & 50 & 25.61 & 24.13 & 25.70 & \underline{25.75} (\textbf{25.81}) & 0.686 & 0.625 & 0.687 & \underline{0.696} (\textbf{0.708}) & 0.363 & 0.443 & 0.313 & \textbf{0.262} (\underline{0.282}) \\
                        \cmidrule(lr){1-2} \cmidrule(lr){3-6} \cmidrule(lr){7-10} \cmidrule(lr){11-14}
                        
\multirow{3}{*}{Set12}  & 15 & \textbf{32.36} & 30.12 & 32.07 & 32.20 (\underline{32.26}) & \textbf{0.895} & 0.837 & 0.889 & 0.891 (\underline{0.894}) & 0.117 & 0.132 & 0.139 & \textbf{0.084} (\underline{0.092}) \\
                        & 25 & \underline{29.93} & 27.54 & \textbf{30.02} & 29.79 (29.76) & \textbf{0.850} & 0.776 & \underline{0.849} & 0.844 (0.848) & 0.159 & 0.218 & 0.159 & \textbf{0.122} (\underline{0.137}) \\
                        & 50 & \textbf{26.71} & 24.67 & 26.49 & \underline{26.60} (26.47) & \textbf{0.768} & 0.683 & 0.734 & 0.755 (\underline{0.760}) & 0.262 & 0.361 & 0.232 & \textbf{0.208} (\underline{0.228})
                        \\ \bottomrule
\end{tabular}
}
\vspace{-1em}
\caption{\textbf{Comparison to the state of the arts on single-image denoising algorithm.} ($\uparrow$) denotes that higher is better and ($\downarrow$) denotes the lower is better. Best performance is in bold. Second best is underlined. For DIP, we report the peak PSNR scores during the optimization.}
\vspace{-1em}
\label{tab:main_results}
\end{table*}

\paragraph{Comparison to DIP variants.}
Table \ref{tab:dips} shows the denoising results of several DIP based methods. 
Deep decoder (DD)~\cite{Deep_heckel_2018_ICLR} shows the worst performance in all metrics.
We believe that DD mitigates overfitting problem with under-parameterized network in return for its performance.
GP-DIP~\cite{GPDIP_Cheng_2019_CVPR} outperforms DIP in PSNR and SSIM.
It uses SGLD~\cite{SGLD_19} to sample multiple instances of posterior distribution and average them which is similar to Self2Self\cite{S2S_Quan_2020_CVPR}.
This strategy may be useful for PSNR score but it may lose the texture of images, which leads to relatively low LPIPS score (see next section for more discussions).
DIP-RED shows the best result apart from our method and its ablated version.
Its plug-and-play overfitting prevention uses other denoising method as prior.
Plug-and-play method might work with our method but it is beyond the scope of this paper.
Note that all above methods except ours, DIP-SURE* and DIP stops optimization at the predefined number of iterations provided in the authors' codes.

In particular, both DIP-SURE*~\cite{DIPSURE_metzler2020unsupervised} and `Ours w/o STE'~\cite{eSURE2019Neurips} are worse than ours even though they use SURE formulation.
We argue that it is because they use a single noise realization.
In addition, they are quite similar each other except DIP-SURE* depends on the $\epsilon$ as a hyper-parameter while `Ours w/o STE' does not have such hyperparameter (Sec.~\ref{sec:Approach}). 
The no need of hyperparameter tuning results in a noticeable gain by `Ours w/o STE.'
Note that original DIP-SURE depends on early stopping by monitoring PSNR with a clean image. 
For fair comparison, we use our stopping criterion to it and notate it as DIP-SURE*.

\vspace{-1em}\paragraph{Comparison to the state of the arts.}
\tablename~\ref{tab:main_results} shows comparative results with other single image denoisning methods in six datasets (four color and two gray-scale).
The comparing methods includes CBM3D~\cite{BM3D_2007_Dabov}, DIP~\cite{DIP_2018_CVPR}, Self2Self (S2S)~\cite{S2S_Quan_2020_CVPR}.
Except for BM3D, all remaining methods are based on convolutional neural network.
For the network architecture for DIP, we use the same one to ours for fair comparison.
We use slightly difference network for S2S since it needs the dropouts instead of batch normalization.

Our method outperforms all other single-image denoising methods in LPIPS, showing comparable PSNR and SSIM.
Ours* exhibits best PSNR performance outperforming all comparing methods except S2S in high noise experiments.
But Ours* loses some high frequency details to ours (see LPIPS).
We believe that it is due to the exponential moving average (EMA) as it alleviates the instability of training (\ie, rough solution space) that cannot be caught by PSNR.
Ours performs well especially at low noise.
We believe that the error by the MC estimation is smaller in the small noise set-up.
Nevertheless, our method exhibits excellent performance in LPIPS and SSIM in almost all setups.


\begin{figure*}[!t]
\centering
    \begin{subfigure}{.15\textwidth}
          \centering 
          \includegraphics[width=1\linewidth]{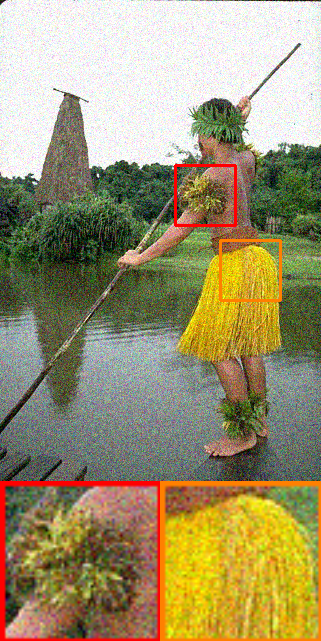}
          \caption*{\centering $\sigma=25$ \break 20.95/0.475}
        \end{subfigure} \hspace{0.1mm}
    \begin{subfigure}{.15\textwidth}
          \centering
          \includegraphics[width=1\linewidth]{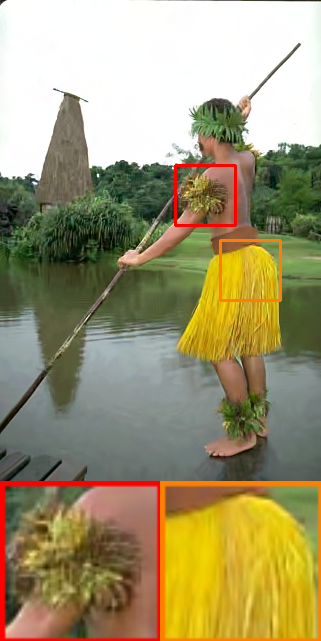}
          \caption*{\centering CBM3D \break \textbf{31.74}/0.096}
        \end{subfigure} \hspace{0.1mm}
    \begin{subfigure}{.15\textwidth}
          \centering
          \includegraphics[width=1\linewidth]{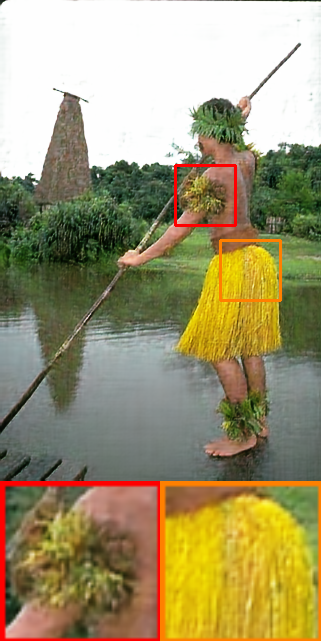}
          \caption*{\centering DIP* \break 29.46/0.095}
        \end{subfigure} \hspace{0.1mm}
    \begin{subfigure}{.15\textwidth}
          \centering
          \includegraphics[width=1\linewidth]{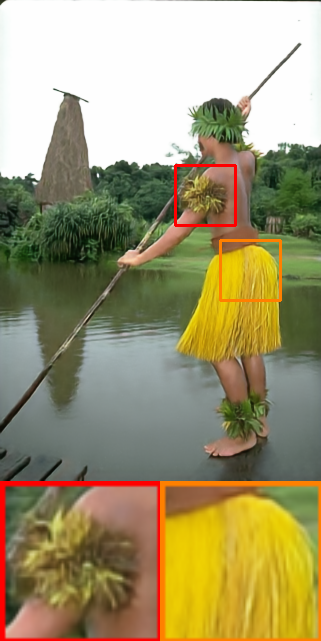}
          \caption*{\centering S2S \break 31.59/\underline{0.088}}
        \end{subfigure} \hspace{0.1mm}
    \begin{subfigure}{.15\textwidth}
          \centering
          \includegraphics[width=1\linewidth]{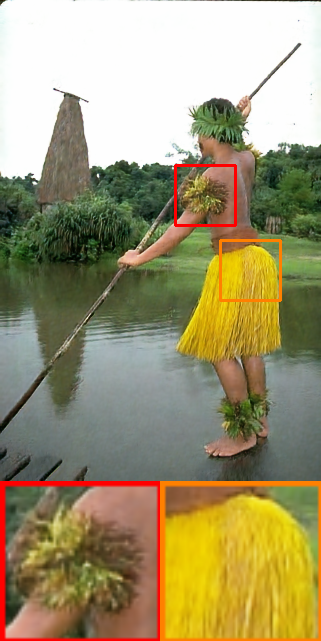}
          \caption*{\centering Ours \break \underline{31.61}/\textbf{0.061}}
        \end{subfigure} \hspace{0.1mm}
    \begin{subfigure}{.15\textwidth}
          \centering
          \includegraphics[width=1\linewidth]{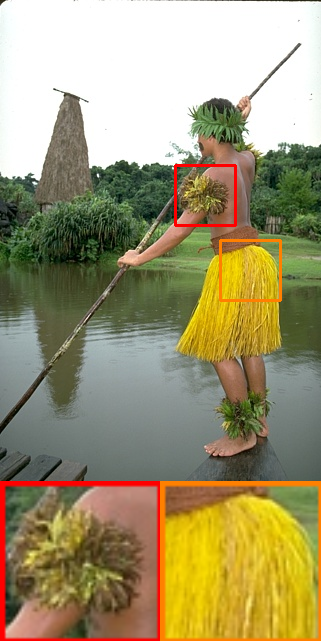}
          \caption*{\centering GT \break PSNR/LPIPS}
        \end{subfigure} \hspace{0.1mm}

\centering
    \begin{subfigure}{.15\textwidth}
          \centering
          \includegraphics[width=1\linewidth]{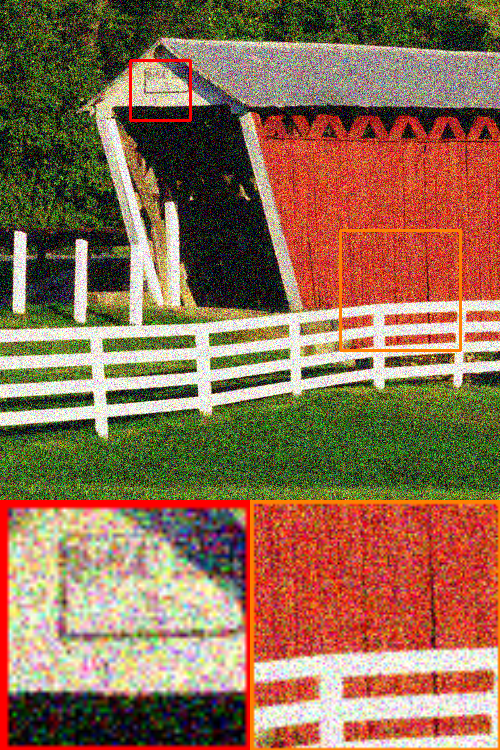}
          \caption*{\centering $\sigma=50$ \break 15.52/0.608}
        \end{subfigure} \hspace{0.1mm}
    \begin{subfigure}{.15\textwidth}
          \centering
          \includegraphics[width=1\linewidth]{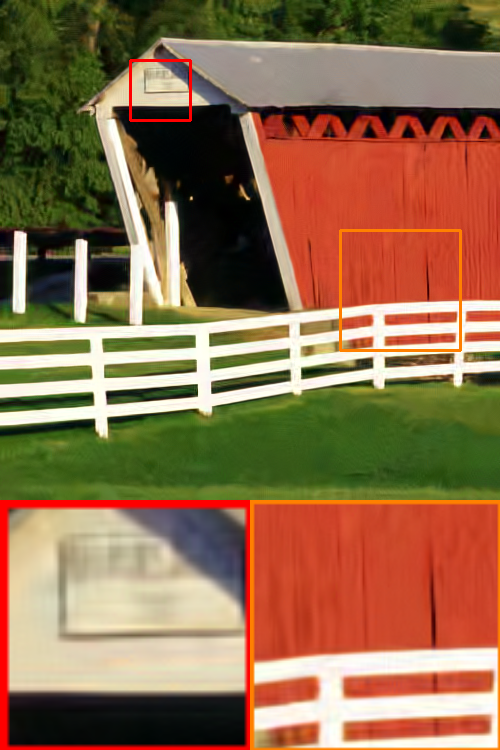}
          \caption*{\centering CBM3D \break 27.82/0.245}
        \end{subfigure} \hspace{0.1mm}
    \begin{subfigure}{.15\textwidth}
          \centering
          \includegraphics[width=1\linewidth]{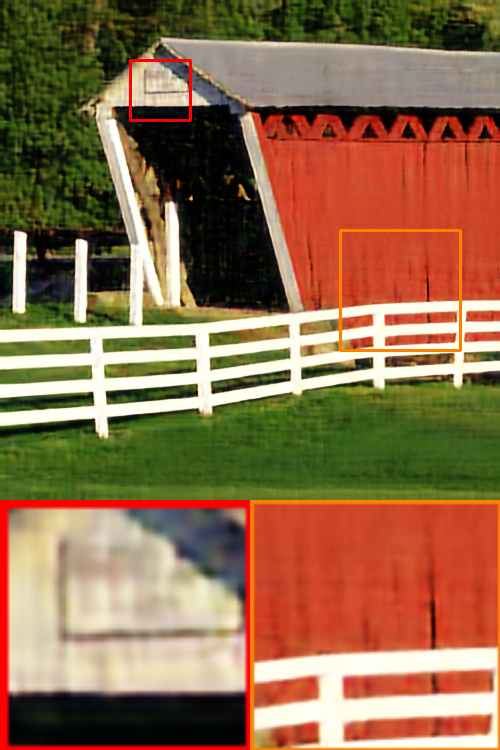}
          \caption*{\centering DIP* \break 26.77/0.265}
        \end{subfigure} \hspace{0.1mm}
    \begin{subfigure}{.15\textwidth}
          \centering
          \includegraphics[width=1\linewidth]{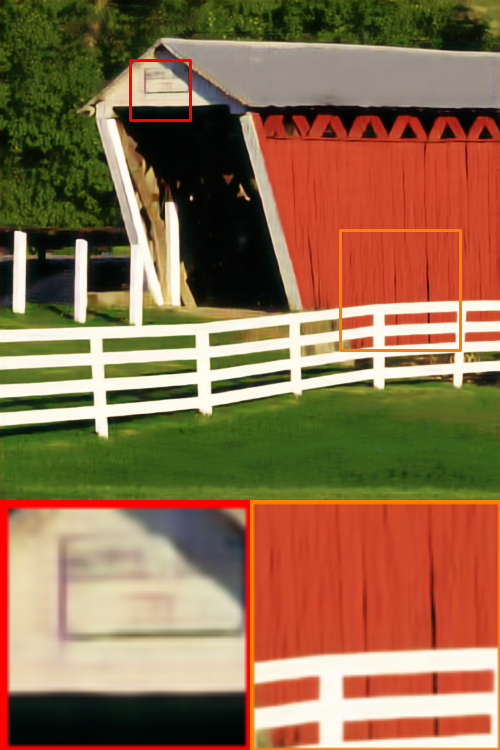}
          \caption*{\centering S2S \break \textbf{28.67}/\underline{0.193}}
        \end{subfigure} \hspace{0.1mm}
    \begin{subfigure}{.15\textwidth}
          \centering
          \includegraphics[width=1\linewidth]{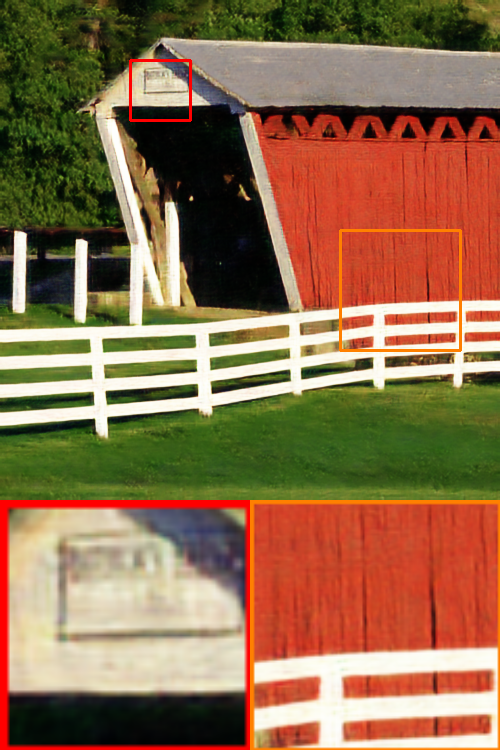}
          \caption*{\centering Ours \break \underline{28.07}/\textbf{0.136}}
        \end{subfigure} \hspace{0.1mm}
    \begin{subfigure}{.15\textwidth}
          \centering
          \includegraphics[width=1\linewidth]{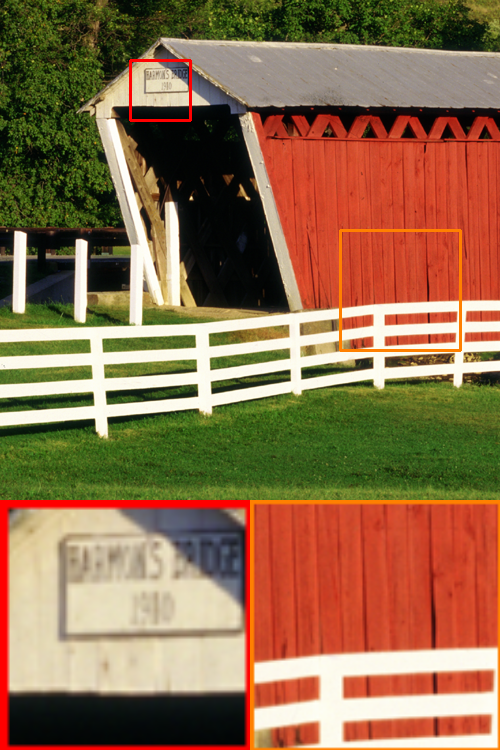}
          \caption*{\centering GT\break PSNR/LPIPS}
        \end{subfigure} \hspace{0.1mm}
    \vspace{-0.2cm}
    \caption{\textbf{Qualitative comparisons.} Best performance: bold. Second best: underlined. More results are in the supplement.}
    \vspace{-1em}
\label{fig:quali}
\end{figure*}

It is worth noting that S2S exhibits high LPIPS, especially in low noise ($\sigma=15$) than all other methods despite being ahead of DIP in PSNR.
Considering that MSE is the sum of squared bias and variance, we argue that the S2S achieves impressive PSNR results with significantly reduced variance and increased squared bias (\ie, destroying textural details).
It is clearly observed in Fig.~\ref{fig:cmp_image_T1}-(b), where we show the results of S2S with various number of ensembles.
As the number of ensembles increases, the PSNR also increases in return of the loss of LPIPS score.
In contrast, our method achieves much better trade-off between LPIPS score and PSNR without ensembling.

Moreover, inference time of S2S on CSet9 is almost 35 hours without parallel processing whereas ours only takes 4 hours.
Further speed up of S2S and ours are possible by parallel processing~\cite{S2S_Quan_2020_CVPR} but the gap would be maintained.

Although it is not quite fair to compare our method with learning-based ones including DnCNN~\cite{DnCNN_2017_TIP}, N2N~\cite{N2N_2018_ICML}, HQ-N2V~\cite{HQN2V2019Neurips}, IRCNN~\cite{IRCNNzhang_2017_learning} as we only use a single noisy observation, we additionally compare with them in the supplementary material for the space sake.

\subsection{Qualitative analysis}
\label{sec:exp_quali}
We present examples of denoised images in Fig.~\ref{fig:quali}.
In the first row, we observe that the results of CBM3D and S2S are over smoothed (having less high frequency details) than those by our method.
DIP preserves textures but is much noisier than ours.
Again, we observe that our results are in better trade-off between PSNR and LPIPS.

The second rows has higher noise level ($\sigma=50$) than the first row.
S2S and CBM3D show clean images with sharp edges.
But they also make the English characters in the sign blurry.
In contrast, our method preserves sharper details in the character in the sign while noises are mostly suppressed.
More qualitative results are in supplement.

\begin{table}[t!]
\resizebox{0.98\linewidth}{!}{
\begin{tabular}{cccca}
\toprule
Noise scale & BM3D-VST~\cite{VST2013Makitalo} & DIP~\cite{DIP_2020_IJCV} & S2S~\cite{S2S_Quan_2020_CVPR} & Ours (Ours*) \\ 
\midrule
$\zeta=0.01$        &   30.50       &  30.99   & \textbf{32.18}     &    \underline{32.00}  (31.94)    \\ 
$\zeta=0.1$         &   21.57       &  23.54   & 22.84     &    \underline{24.87}  (\textbf{24.94}) \\
$\zeta=0.2$         &   18.48       &  21.43   & 20.10     &    \underline{22.85}  (\textbf{22.90})    \\ 
\bottomrule
\end{tabular}
}
\caption{\textbf{Comparison to the state of the arts on Poisson noise (PSNR (dB)).} Best performance: bold. Second best: underlined.}
\vspace{-1.5em}
\label{tab:Poisson}
\end{table}

\subsection{Extension to Poisson noise}
\label{sec:poisson_exp}
Poisson noise is likely to occur in low light condition such as microscopic imaging.
In~\cite{PURE2018Soltanayev}, they use MNIST images for simulating this scenario.
We conduct experiments of single-image Poisson denoising, and summarize the comparative results with BM3D-VST~\cite{VST2013Makitalo}, DIP, and S2S in Table~\ref{tab:Poisson}.
Note that BM3D-VST is one of the most popular methods for Poisson denoising.

For low noise level ($\zeta =0.01$), noise distribution becomes almost symmetric similar to Gaussian.
So, our method does not perform well.
But at higher level of noise, our method outperforms other methods.
DIP shows better results than classic methods such as BM3D with VST~\cite{VST2013Makitalo}, and the state of the art, S2S, in the higher noise.
Our method outperforms all compared methods including the classic methods with VST (BM3D-VST)~\cite{VST2013Makitalo}.

\begin{figure}[t!]
\centering
    \begin{subfigure}{.09\textwidth}
      \centering
      \includegraphics[width=1\linewidth]{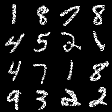}
      \caption*{\centering \scriptsize Noise \break 18.81/0.053} 
    \end{subfigure} \hspace{0.1mm} 
    \begin{subfigure}{.09\textwidth}
      \centering
      \includegraphics[width=1\linewidth]{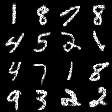}
      \caption*{\centering\scriptsize BM3D-VST~\cite{BM3D_2007_Dabov}\break\scriptsize{19.87/0.040}} 
    \end{subfigure} 
    \begin{subfigure}{.09\textwidth}
      \centering
      \includegraphics[width=1\linewidth]{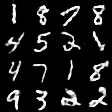}
      \caption*{\centering\scriptsize DIP~\cite{DIP_2020_IJCV} \break \underline{21.17} / \underline{0.024}} 
    \end{subfigure} 
    \begin{subfigure}{.09\textwidth}
      \centering
      \includegraphics[width=1\linewidth]{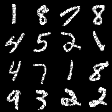}
      \caption*{\centering\scriptsize S2S~\cite{S2S_Quan_2020_CVPR} \break 20.09 / 0.040} 
    \end{subfigure} 
    \begin{subfigure}{.09\textwidth}
      \centering
      \includegraphics[width=1\linewidth]{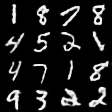}
      \caption*{\centering\scriptsize Ours \break \textbf{22.99} / \textbf{0.020}} 
    \end{subfigure} \hspace{0.1mm} 
\vspace{-1em}
\caption{\textbf{Qualitative comparison on Poisson noise ($\zeta=0.2$).}}
\label{fig:quali_Poisson}
\vspace{-2em}
\end{figure}

Fig.~\ref{fig:quali_Poisson} shows the qualitative result of the Poisson noise setup.
We observe that BM3D+VST images were considerably blurrier than other methods, and S2S also produce blurry image due to overfitting.
DIP shows the second best result thanks to early stopping.
In contrast, our method denoises holes in the images with detailed texture preserved without early stopping.

\section{Conclusion}
We investigate DIP for denoising by the notion of effective degrees of freedom to monitor the overfitting to noise and propose stochastic temporal ensembling (STE) and zero crossing stopping criterion to stop the optimization before it overfits without a clean image.
We significantly improve the performance of Gaussian denoising by DIP without the manual early stopping and extend the method to Poisson denoising with PURE.
Our empirical validation shows that the proposed method outperforms state-of-the-arts in LPIPS by large margins with comparable PSNR and SSIM, evaluated with the seven different datasets.

\vspace{0.5em}
{
\footnotesize
\noindent
\textbf{Acknowledgement.} This work was partly supported by the National Research Foundation of Korea (NRF) grant funded by the Korea government (MSIT) (No.2019R1C1C1009283) and Institute of Information \& communications Technology Planning \& Evaluation (IITP) grant funded by the Korea government (MSIT) (No.2019-0-01842, Artificial Intelligence Graduate School Program (GIST)), (No.2019-0-01351, Development of Ultra Low-Power Mobile Deep Learning Semiconductor With Compression/Decompression of Activation/Kernel Data, 17\%), (No. 2021-0-02068, Artificial Intelligence Innovation Hub) and was conducted by Center for Applied Research in Artificial Intelligence (CARAI) grant funded by DAPA and ADD (UD190031RD). The work of SY Chun was supported by Basic Science Research Program through National Research Foundation of Korea (NRF) funded by Ministry of Education (NRF-2017R1D1A1B05035810). \par
}

{\small
\bibliographystyle{ieee_fullname}
\bibliography{egbib}
}
\end{document}